\documentclass[12pt]{iopart}
\usepackage{epsfig,graphicx,amssymb}

\begin{document}

\title[]{The Geometrical Effects on Electronic Spectrum and Persistent
Currents in Mesoscopic Polygon}

\author{Shengli Zhang, Qi Wang, Erhu Zhang}

\address{Department of Applied Physics, Xi'an Jiaotong University, \\
Xi'an 710049, People's Republic of China}

\begin{abstract}
In this paper, a new mesoscopic polygon which possesses smooth
transition at its corners is proposed. Because of the particularity
of structure, this kind of mesoscopic polygon can also be a
geometrical supperlattice. The geometrical effects on the electron
states and persistent current are investigated comprehensively in
the presence of magnetic flux. We find that the particular geometric
structure of the polygon induces an effective periodic potential
which results in gaps in the energy spectrum. The changes of gaps
show the consistency with the geometrical twoness of this new
polygon. This electronic structure and the corresponding physical
properties are found to be periodic with period $\phi_{0}$ in the
magnetic flux $\phi $ and can be controlled by the geometric method.
We also consider the Rahsba spin-orbit interaction which make the
energy levels splitting newly to double and leads to an additional
small zigzag in one period of the persistent current. These new
phenomena may be useful for the applications in quantum device
design in the future.

\end{abstract}

\maketitle

The rapid developments of advanced growth techniques make it
possible to fabricate reduced-dimensional quantum systems with
complex geometries which attract much attentions in recent
years$^{[1-4]}$. The fabrication of essentially arbitrary geometries
could lead to dramatic control of the electronic properties of
solids by means of geometries. For example, ringlike structures of
semiconductor have become the subject of extensive theoretical and
experimental studies because of their unique topologies and
potential applications in the spintronics nanodevices$^{[5]}$, which
utilize the spin rather than the charge of an electron. A prime
candidate for spin manipulation is the Rashba spin-orbit interaction
(SOI)$^{[6]}$, which stems from the absence of structure inversion
symmetry. For ring structures, the energy spectrum and the
interference effects due to the SOI have been well
investigated$^{[7]}$ and for square loops it has been shown$^{[8]}$
that interference due to Rashba SOI can lead to electron
localization. Recently, the spin-interference of ballistic electrons
traveling along any regular polygon is also studied$^{[9,10]}$ and
shows a dependence on the sidelength and alignment of the polygon as
well as on the SOI constant.

\begin{figure}[!b]
\centering
\includegraphics[angle=270,scale=0.25]{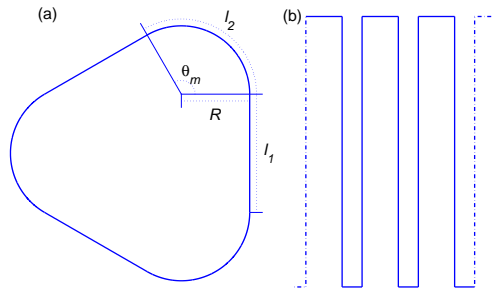}
\caption{(a) Sketch of a mesoscopic polygon ($N=3$) in the x-y
plane. (b) The corresponding periodic structures of the geometrical
potential.}
\end{figure}

Besides the quantum rings, many of the fabrication structures
exhibit curvature on the nanoscale, such as nanotubes$^{[2]}$,
nanotori$^{[3]}$, the spiral inductors$^{[11]}$, quantum snakes, and
so on. These curvilinear structures have been the focus on
investigation of new physical features that appear due to the
curvature.$^{[12]}$ When the electron is strongly confined to a low
dimensional curvilinear system with smooth geometry, the interplay
between geometry and quantum physics results in an effective
geometric potential whose magnitude depends on the local
curvature.$^{[13]}$ This implies that the quantum behaviors of the
electron can be controlled by altering the local geometric
curvature. Such an effective geometric potential has successfully
applied, for example, to the band-structure calculation of real
systems$^{[14]}$, to the determination of electron states in
curvilinear quantum wires$^{[15]}$ and to the impact of curvature on
electron transportation in different curved systems$^{[16,17,18]}$.
Furthermore, experiment of localized states of high oxidized porous
silicon$^{[19]}$ indicates that the curvature potential plays
important roles in the electronic behaviors. However, the studies on
electronics and spintronics of the ringlike structure, square loop
and regular polygon did not deal with the curvature potential. What
are the effects of the curvature potential is significant to
investigate in curved mesoscopic systems.

In this paper, we focus on the polygonal structures of the semiconductor$%
^{[20]}$. Considering that the edges of the nano-polygon always have
the width, we establish an extended one-dimensional model of
mesoscopic regular polygon with smooth transition at its corners. We
point out that this mesoscopic polygon takes more further freedom on
controlling the electron behaviors. The curvature effects are taken
into account through introducing the geometrical potential. Because
magnetic field effects prove important to the interference
phenomena, the electronic spectrum and persistent currents of
electrons in the mesoscopic polygon are theoretically investigated
in the presence of magnetic flux. The Rashba SOI is also considered.

Geometrically, our mesoscopic polygon consists of $N$ straight line
segments and $N$ arc segments forming a periodic structure with the
$C_{N}$ symmetry. As shown in Fig. 1 (a), the circumference is
$L=N\left( l_{1}+l_{2}\right) $ where $l_{1} $ is the length of one
straight line and $l_{2}=R\theta
_{m}=2\pi R/N$ for an arc with the curvature radius $R$\ and the arc angle $%
\theta _{m}$. We can see that this mesoscopic polygon, which trends
to a perfect ring for the limit of $l_{1}\rightarrow 0$ and to a
regular polygon for the limit of $R\rightarrow 0 $, shows a
geometrical twoness. The curvature can arise an effective geometric
potential with the form $V_{g}=-\hbar ^{2}\kappa ^{2}/(8m)^{[17]}$,
where $\kappa $ is the curvature of the wire. Form Fig. 1 (b), it is
obvious that the geometric potentials form a structure of periodic
square potential. So the mesoscopic polygon can be regarded as a
geometrical supperlattice.

We consider a mesoscopic polygon in the $x-y$ plane subjected to an
axial magnetic field $B $ which is oriented through the $z$ axis.
For a convenient
choice of magnetic vector, the Hamiltonians in the presence of the Rashba SOI%
$^{[6]}$ for a conduction electron of the effective mass $m$ are given by$%
^{[18]}$%
\begin{eqnarray}
H_{line}=-\frac{\hbar ^{2}}{2m}\frac{\partial ^{2}}{\partial s^{2}}+i%
\frac{e\hbar }{c}A_{s}\frac{\partial }{\partial s}+\frac{e^{2}}{2mc^{2}}%
A_{s}^{2}-i\alpha \sigma _{b}(\frac{\partial }{\partial s}+i\frac{e\hbar }{c}A_{s}),\\%
H_{arc}=-\frac{\hbar ^{2}}{2m}\frac{\partial ^{2}}{\partial s^{2}}+i\frac{%
e\hbar }{c}A_{s}\frac{\partial }{\partial s}-\frac{\hbar ^{2}\kappa ^{2}}{8m}%
+\frac{e^{2}}{2mc^{2}}A_{s}^{2}-i\alpha \lbrack \sigma_{b}(\frac{\partial }{%
\partial s}+i\frac{e\hbar }{c}A_{s})-\frac{1}{2}\sigma_{t}\kappa],
\end{eqnarray}%
where $A_{s}$ is the component of vector potential ${\bf A}$\ along
the arclength $s$
direction, $\sigma _{b}={\bf \hat{\sigma}}\cdot $ ${\bf b}$ and $\sigma _{t}=%
{\bf \hat{\sigma}}\cdot $ ${\bf t}$, expressed by the usual Pauli matrices $%
\hat{\sigma}_{x,y,z}$, are the spin matrices on the normal ${\bf b}$
and tangent ${\bf t}$ direction, respectively. The parameter $\alpha
$ is the Rashba strength which represents the average electric field
along the $z$ direction and can be controlled by a gate voltage.

Both the wave function in the straight lines $\Psi _{1}$ and the
wave function in the arc $\Psi _{2}$ should satisfy the Bloch
theorem and two boundary conditions$^{[21]}$:

(i) the Bloch theorem reads
\begin{equation}
\Psi _{1,2}(s+l_{1}+l_{2})=e^{iK(l_{1}+l_{2})}\Psi _{1,2}(s),
\end{equation}%
where $K$ is the wave vector in the reciprocal-space.

(ii) the boundary conditions at $s=0,L$ ($L=N(l_{1}+l_{2})$ is the
circumference of mesoscopic polygon) are
\begin{equation}
\Psi _{1}(0)=e^{i2\pi (\phi /\phi _{0})}\Psi _{2}(L),\frac{\partial
\Psi _{1}(0)}{\partial s}=\frac{\partial \Psi _{2}(L)}{\partial s},
\end{equation}%
where $\phi $ is the magnetic flux along the $z$ direction through
the area confined by the mesoscopic polygon and $\phi _{0}$ is the
magnetic flux quantum. From Eqs. (3) and (4), we get the
relationship $KL=2\pi(n+\phi /\phi _{0}), n=0, 1, 2, ...$. It is
obviously that Eq. (4) imply that the electronic spectra of
mesoscopic polygon are periodic in $\phi$ with period $\phi_{0} $,
and so are the other physical properties.

(iii) the boundary conditions at the connecting points $%
s=m(l_{1}+l_{2})+l_{1},m=0,1,...,N-1$ or$\
s=m(l_{1}+l_{2}),m=1,...,N-1$ are

\begin{equation}
\Psi _{1}(s)=\Psi _{2}(s),\frac{\partial \Psi _{1}(s)}{\partial s}=\frac{%
\partial \Psi _{2}(s)}{\partial s}.
\end{equation}

\begin{figure}[!b]
\centering
\includegraphics[scale=0.3]{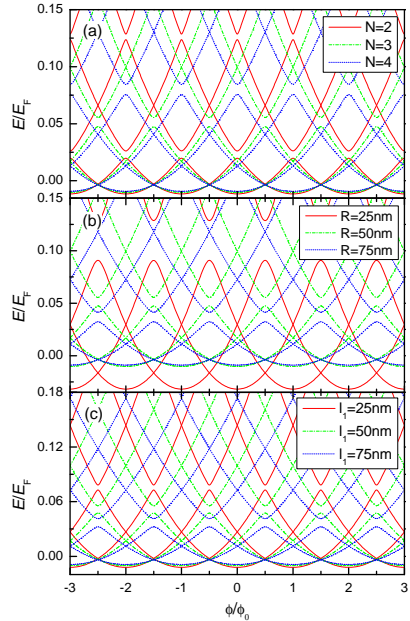}
\caption{The electron energy levels in the mesoscopic polygon
as a function of the magmatic flux $\phi $ in the absence of SOI ($\alpha =0$%
). The energy spectrum is shown for $N=2$, $3$, $4$ at $R=50$ nm and
$l_{1}=$ $50$ nm in (a), for $R=25$ nm, $50$ nm, $70$ nm at $N=3$
and $l_{1}=$ $50$ nm in (b), and for $l_{1}=25$ nm, $50$ nm, $75$ nm
at $N=3$ and $R=$ $50$ nm in (c).}
\end{figure}

When there isn't Rashba SOI, i. e. $\alpha =0$, using the above
conditions we can easily obtain a transcendental equation defining
the energy spectrum
for unbound states $E>0$%
\begin{equation}
\cos (k_{2}l_{2})\cos (k_{1}l_{1})-\sin (k_{2}l_{2})\sin
(k_{1}l_{1})\left( \frac{k_{1}^{2}+k_{2}^{2}}{2k_{1}k_{2}}\right)
=\cos[\frac{2\pi}{N}(n+\frac{\phi}{\phi_{0}})]
\end{equation}%
where $k_{1}=\sqrt{2mE/\hbar ^{2}}$ is the wave vector for straight
lines and $k_{2}=\sqrt{2m(E+V_{g})/\hbar ^{2}}$ for arc. While for
bounded states $-V_{g}<E<0$%
\begin{equation}
\cos (k_{2}l_{2})\cosh (\tilde{k}_{1}l_{1})-\sin (k_{2}l_{2})\sinh (\tilde{k}%
_{1}l_{1})\left( \frac{\tilde{k}_{1}^{2}-k_{2}^{2}}{2\tilde{k}_{1}k_{2}}%
\right) =\cos[\frac{2\pi}{N}(n+\frac{\phi}{\phi_{0}})],
\end{equation}%
here $\tilde{k}_{1}=\sqrt{-2mE/\hbar ^{2}}$ and $\phi $ is the
magnetic flux through the area confined by the mesoscopic polygon.

For InAs material, the effective electron mass $m=0.023m_{0}$ and
the Fermi energy $E_{F}=11.13\times10^{-3}$ eV. The calculated
energy spectra as a function of the magnetic field flux for a
different set of $N$, $R$ and $l_{1}$ are given in Fig. 2,
respectively. Different from the status in the perfect rings where
the energy levels are intersecting parabolas, the gaps are opened at
the points of intersection of the parabolas in our mesoscopic
polygon. The whole energy levels are periodic curves with a period
of $\phi _{0}$. In addition, if the mesoscopic polygon circumference
increases through changing the geometrical parameters $N,$ $R$ and
$l_{1}$, the energy levels tend to flat and have the negative shifts
which are larger at higher energy. It is obvious that there exit the
bound states of $E<0$ in the energy spectra. Our results are
consistent with the known fact that there is one and only one bound
state for $\theta _{m}=2\pi /N\leq \pi $ in such a quadrate trap
formed by the geometric potential$^{[12]}$. Additionally, one can
see that the bound state of the mesoscopic polygon can transform to
the unbound state smoothly in the presence of the magnetic flux. So
there will be a contribution to the persistent current from the
bound state.

When there is Rashba SOI, the transcendental equation defining the
energy spectrum for unbound states $E>0$ obtained from the boundary
conditions is

\begin{eqnarray}
&&-2f^{2}\{\cos [\frac{4\pi }{N}(\frac{\phi }{\phi
_{0}}+n+\frac{1}{2})]+\cos(2bl_{1})+\cos (2dl_{2})+3\cos (dl_{2})\cos (bl_{1})  \nonumber \\
&&-4\cos (2\lambda )\sin ^{2}(dl_{2})\sin ^{2}(bl_{1})-4\sin
(\lambda )\sin
(2dl_{2})\sin (2bl_{1})\}  \nonumber \\
&&-16f\cos [\frac{2\pi }{N}(\frac{\phi }{\phi
_{0}}+n+\frac{1}{2})][\cos(dl_{2})\cos(bl_{1})-\sin(dl_{2})\sin(bl_{1})\sin
(\lambda )] \nonumber
\\
&&\times \lbrack 2f\cos (cl_{2})\cos (al_{1})-(1+f^{2})\sin
(cl_{2})\sin
(al_{1})]  \nonumber \\
&&-[(1+6f^{2}+f^{4})\cos (2cl_{2})-(1-f^{2})^{2}]\cos (2al_{1})  \nonumber \\
&&+4f(1+f^{2})\sin (2cl_{2})\sin (2al_{1})+(1-f^{2})^{2}\cos
(2cl_{2})
\nonumber \\
&=&1+4f^{2}+f^{4},
\end{eqnarray}%
For bound states $-V_{g}<E<0$, the transcendental equation in the
presence of Rashba SOI is
\begin{eqnarray}
&&2\tilde{f}^{2}\{4\cos [\frac{4\pi }{N}(\frac{\phi }{\phi _{0}}+n+\frac{1}{2}%
)]+\cos (2bl_{1})+\cos (2dl_{2})+3\cos (dl_{2})\cos (bl_{1})  \nonumber \\
&&-4\cos (2\lambda )\sin ^{2}(dl_{2})\sin ^{2}(bl_{1})-4\sin
(\lambda )\sin
(2dl_{2})\sin (2bl_{1})\}  \nonumber \\
&&-16\tilde{f}\cos [\frac{2\pi }{N}(\frac{\phi }{\phi _{0}}+n+\frac{1}{2}%
)][\cos (dl_{2})\cos (bl_{1})-\sin (dl_{2})\sin (bl_{1})\sin
(\lambda )]
\nonumber \\
&&\times \lbrack 2\tilde{f}\cos (cl_{2})\cosh (\tilde{a}l_{1})+(\tilde{f}%
^{2}-1)\sin (cl_{2})\sinh (\tilde{a}l_{1})]  \nonumber \\
&&-[(1-6\tilde{f}^{2}+\tilde{f}^{4})\cos (2cl_{2})-(1+\tilde{f}%
^{2})^{2}][\sinh ^{2}(\tilde{a}l_{1})+\cosh ^{2}(\tilde{a}l_{1})]
\nonumber
\\
&&+4\tilde{f}(\tilde{f}^{2}-1)\sin (2cl_{2})\sinh (2\tilde{a}l_{1})+(1+%
\tilde{f}^{2})^{2}\cos (2cl_{2})  \nonumber \\
&=&1-4\tilde{f}^{2}+\tilde{f}^{4},
\end{eqnarray}%
where the angle $\lambda $ is given by $\tan \lambda =-\Delta $ with $%
\Delta=2m\alpha R/\hbar ^{2}$, $a=\Delta _{1}/(2R)$, $b=\Delta /(2R)$, $%
c=\Delta_{2}/(2R)$, $d=\Delta _{3}/(2R)$, $\tilde{a}=\sqrt{-\Delta ^{2}-%
\frac{E}{V_{{\normalsize g}}}}/(2R)$, $\Delta _{1}=\sqrt{\Delta ^{2}+\frac{E%
}{V_{{\normalsize g}}}}$, $\Delta _{2}=\sqrt{\Delta ^{2}+\frac{E}{V_{%
{\normalsize g}}}+1}$ ,$\Delta _{3}=\sqrt{\Delta ^{2}+1}$, $f=a/c$ and $%
\tilde{f}=\tilde{a}/c$.

The corresponding energy levels as a function of the magnetic flux
are shown
in Fig. 3 for different $N$, $R$, and $l_{1}$, respectively, when $%
\alpha=1.0\times 10^{-11}$ eVm. Due to the spin-orbit interaction,
all of the energy levels split to double corresponding to spin-up
and spin-down. Besides this, the other characters of the energy
spectra is similar with the results in the absence of SOI. For
instance, the sub-energy levels take the period of $N\phi _{0}$ and
the gaps also appear at the intersection points of the parabolas.

\begin{figure}[!h]
\centering
\includegraphics[scale=0.3]{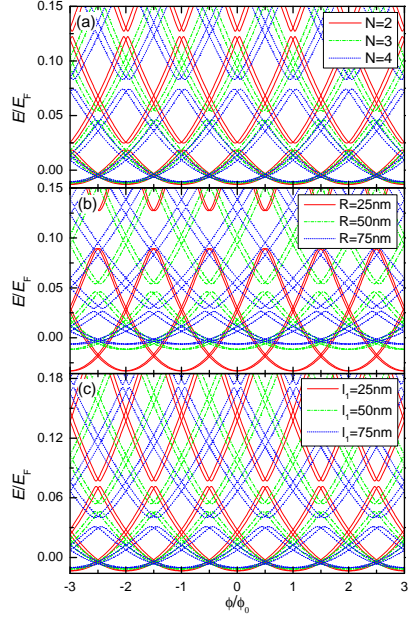}
\caption{The electron energy levels in the mesoscopic polygon as a
function of the magmatic flux $\phi $ in the presence of SOI with $
\alpha =1.0\times 10^{-11}$ eVm. The parameters are the same sa Fig.
2.}
\end{figure}

The energy gaps $E_{{\normalsize g}}$ as functions of $l_{1}$ are
plotted in Fig. 4 for no SOI. Panel (a) is for the energy gaps which
are between the first energy level and the second energy level of
the energy spectra in Fig. 2(c). The energy gaps, which are between
the third energy level and the fourth energy level, are shown in panel (b). Clearly, the gaps $E_{%
{\normalsize g}}$ can be effectively modulated by the geometrical
parameter of the mesoscopic polygon. On the whole, the profile of
the curves is always increase first and then decrease. For the limit
of $l_{1}\rightarrow 0$, the energy gaps is small and decrease to
zero quickly. While for another limit of $l_{1}\rightarrow \infty $,
the energy gaps is also small, which shows the geometrical twoness
of the mesoscopic polygon. For different values of $R $, the curves
of the energy gaps take the rightward shifts which are larger at
large radius. This may be due to the competition between the radius
$R$ and length $l_{1}$. If the radius $R$ increasing, a longer
length of $l_{1}$ is needed to exhibit the property of regular
polygon. In addition, because of the transmission and reflection by
the geometrical potential, the electrons in the mesoscopic polygon
has a more complex interference than that in the perfect ring. This
can be seen from the energy gaps in panel (b) where there exit two
vibrations.

\begin{figure}[!h]
\centering
\includegraphics[scale=0.3]{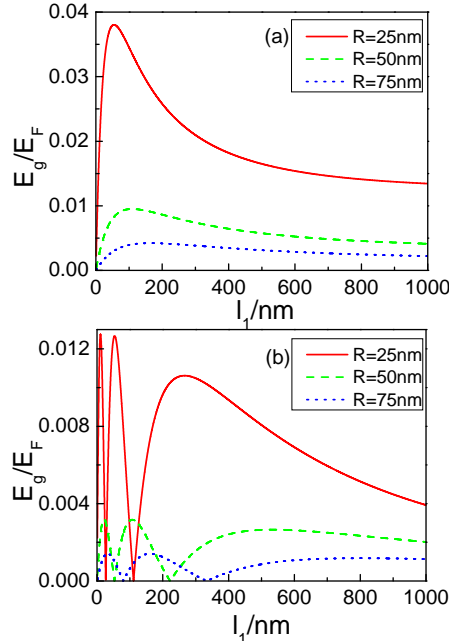}
\caption{The relationship of the energy gaps in the mesoscopic
polygon vs $l_{1}$ for $R=25$ nm, $50$ nm, $70$ nm at $N=3$ in the
absence of SOI. (a) is for the energy gaps which are between the
first energy level and the second energy level, and (b) is for the
energy gaps which are between the third energy level and the fourth
energy level.}
\end{figure}

For the energy gaps changing with the parameters $R$ or $N$, our
calculations show the similar results as above, except for
$R\rightarrow 0$
where the geometrical potential tends to infinite, i. e. $V_{{\normalsize g}%
}\rightarrow -\infty $, and the motions of electrons may be
localized or isolated by these infinite potentials. So the energy
gaps are increased to an finite value. The above-all mentioned
behaviors can also be found for the energy gaps in the presence of
Rashba SOI.

\begin{figure}[!b]
\centering
\includegraphics[scale=0.3]{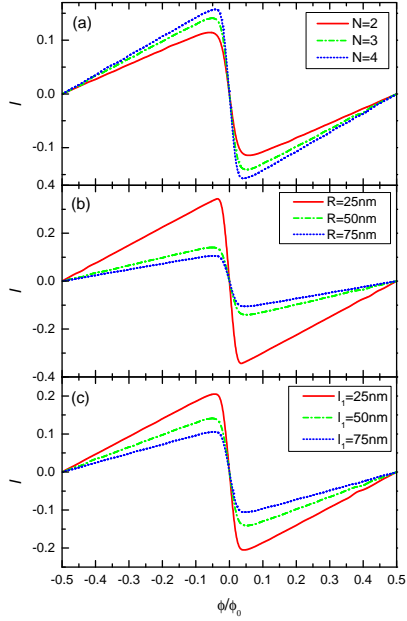}
\caption{Persistent current in the mesoscopic polygon vs. magnetic
flux $\phi $\ at $\alpha =0$: (a) for $N=2$, $3$, $4$ at $R=50$ nm
and $l_{1}=$ $50$ nm, (b) for $R=25$ nm, $50$ nm, $70$ nm at $N=3$
and $ l_{1}=$ $50$ nm, (c) for $l_{1}=25$ nm, $50$ nm, $75$ nm at
$N=3$ and $R=$ $ 50$ nm. The persistent current is in units of
$E_{{\normalsize F}}/\phi _{0}$.}
\end{figure}

Now we proceed to investigate the persistent currents induced by the
magnetic flux. At zero temperature, it is given by$^{[22]}$
\begin{equation}
I=-%
\mathrel{\mathop{\sum }\limits_{i\in {\normalsize occupied}}}%
\frac{\partial E_{i}}{\partial \phi },
\end{equation}%
where $E_{i}$ are the single particle eigenenergies. This formula is
valid only in the absence of electron-electron interactions, which
we neglect here.

For the given mesoscopic polygon with the filled lowest two bands,
the persistent currents in the absence of SOI are calculated from
the energy levels in Fig. 2 and shown in Fig. 5. We have known that
the persistent current oscillations in a perfect ring have a
sawtooth form at zero temperature. However, the mesoscopic polygon
shows a smoothing of the oscillations. The smoothing and the swing
of oscillation are decided not only by the gaps opened at the
intersection points, but also by the geometrical structure of the
mesoscopic polygon. We can see that the amplitude of oscillation,
which is determined by the band width of the energy spectra,
increases as the parameter $N$ increases and decreases as the radius
$R$ and length $l_{1}$ increase.

\begin{figure}[!b]
\centering
\includegraphics[scale=0.3]{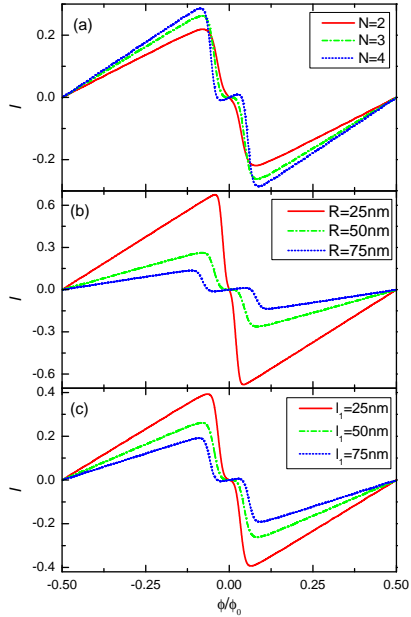}
\caption{Persistent current in the mesoscopic polygon vs. magnetic
flux $\phi $\ in the presence of SOI with $\alpha =1.0\times
10^{-11}$ eVm. The parameters are the same as Fig. 5. The persistent
current is in units of $E_{{\normalsize F}}/\phi _{0}$.}
\end{figure}

While the SOI is considered, the corresponding persistent currents
are shown in Fig. 6 according to the energy spectra in Fig. 3. On
the whole, the oscillation amplitude of persistent current is
similar to the results without the SOI. While particularly, the
oscillation in one period shows an additional small zigzag which is
due to the energy-split induced by SOI seen in Fig. 3. As we
increase the parameter $N$, the whole fluctuation range of the
persistent current increases accordingly. However the whole
fluctuation ranges decrease as the increscences of $R$ and $l_{1}$.
While the increases in $N$, $R$ and $l_{1}$ will all broaden and
enlarge the small zigzag.

In summary, we study the mesoscopic polygon with round corners and
investigate the geometrical effects on the quantum behaviors of a
single electron in it under the influences of magnetic flux.
Different from the studies of other mesoscopic structures, the
geometric potentials which can bring on the new phenomena have been
considered in this new system. We calculate the electron states and
persistent currents changing with magnetic flux $\phi $ under two
circumstances, with SOI and without SOI, respectively. The
geometrical structure results in an effective periodic potential and
leads to a new electronic structure accompanying with the energy
gaps. The changes of the gaps displays the geometrical twoness of
the mesoscopic polygon. The SOI results in the split energy levels
and induces the unique periodic structure for the persistent current
which is formed by a large zigzag and a small zigzag in one period.
We also find that the energy spectrum and the related physical
properties can be modulated by the geometrical methods. It can be
concluded that the rich structures in the mesoscopic polygon can
provide more freedom to tailor the electronic structures and then
effectively control the related properties. So we may construct the
custom-built quantum device from the ring geometry to the polygon
geometry, even build a geometrical supperlattice.

The authors thank Dr S. Zhao, L. Zhang, R. Liang, Y. Liu, and Z. Yao
for helpful discussions. This work is supported by NSF of China
under Grant No. 10374075. E. Zhang is also supported by the Doctoral
Foundation Grant of Xi'an Jiaotong University (XJTU) No.
DFXJTU2004-10 and by the NSF of XJTU (Grant No. 0900-573042).

\end{document}